\documentclass{ws-procs961x669}            
\begin{document}
\title{Accelerated expansion of the Universe and the Higgs true vacuum}

\author{Muhammad Usman$^*$ and Asghar Qadir$^{\dagger}$}

\address{Department of Physics, \\
	School of Natural Sciences (SNS), \\
	National University of Sciences and Technology (NUST), \\
	Sector H-12 Islamabad 44000 Pakistan. \\
	$^*$E-mail: muhammad\_usman\_sharif@yahoo.com \\
	$^{\dagger}$E-mail: asgharqadir46@gmail.com}

\begin{abstract}
Scalar fields which are favorite among the possible candidates for the dark energy usually have degenerate minima at $\pm \phi_{min}$. In the presented work, we discuss a two Higgs doublet model with the non-degenerate vacuum named inert uplifted double well type two-Higgs doublet model (UDW-2HDM) for the dark energy. It is shown that when the both Higgs doublets lie in their respective true minima then one Higgs doublet can cause the current accelerated expansion of the Universe.
\end{abstract}


\bodymatter
\section[Introduction]{Introduction}\label{sec:introduction}
Modeling the accelerated expansion of the Universe which is also in agreement with at least the standard model of the Particle Physics is a very hot question. We in this context earlier showed that the Higgs field(s) of the inert doublet model (IDM) and the general two Higgs doublet model (2HDM) can provide the current accelerated expansion of the Universe. In this short work we show that a new kind of two Higgs doublet model where both the Higgs vacuums are non-degenerate can also be used for dark energy modeling. We call this model to be uplifted double well two Higgs doublet model (UDW-2HDM).
We in this short review give the very brief review and the results with necessary model information. The complete full length paper will be submitted to an ISI indexed journal later.
\section[UDW type 2HDM]{Uplifted double well two-Higgs double model}\label{sec:UDW-2HDM}
The Lagrangian which describes any model in Particle Physics is
\begin{equation}\begin{array}{rcl}\label{L}
\mathcal{L}=\mathcal{L}^{SM}_{gf}+\mathcal{L}_{Y}+T_{H}-V_{H}~,
\end{array}
\end{equation}
where $\mathcal{L}^{SM}_{gf}$ is the $SU_{C}(3){\otimes}SU_{L}(2){\otimes}U_{Y}(1)$ SM interaction of the fermions and gauge bosons (force carriers), $\mathcal{L}_{Y}$ is the Yukawa interaction of fermions with the Higgs field(s), $T_{H}$ is the kinetic term of the Higgs field and $V_{H}$ is the potential of the Higgs field. The last two terms form the Higgs Lagrangian $\mathcal{L}_{Higgs}$. In UDW-2HDM
\begin{equation}\begin{array}{rcl}\label{TH}
T_H= ({D_1}_{\mu}\phi_1)^\dagger ({D_1}^{\mu}\phi_1) &+& ({D_2}_{\mu}\phi_2)^\dagger ({D_2}^{\mu}\phi_2) \\ & & +\left [ {\chi}({D_1}_{\mu}{\phi}_{1})^{\dagger}({D_2}^{\mu}{\phi}_{2})+{\chi^{*}}({D_2}_{\mu}{\phi}_{2})^{\dagger}({D_1}^{\mu}{\phi}_{1}) \right ],
\end{array}
\end{equation}
and
\begin{equation}\begin{array}{rcl}\label{VH}
V_H & = &
\rho_1\exp(\Lambda_1m_{11}^2(\phi_1-\phi_{1_0})^\dagger(\phi_1-\phi_{1_0}))+ \rho_3\exp\left(\dfrac{1}{2}\Lambda_3\lambda_1((\phi_1-\phi_{1_0})^\dagger(\phi_1-\phi_{1_0}))^2\right) 
\\ & & \hspace{-0.25cm}
+\rho_2 \exp(\Lambda_2 m_{22}^2(\phi_2-\phi_{2_0})^\dagger(\phi_2-\phi_{2_0}))+\rho_4\exp\left(\dfrac{1}{2}\Lambda_4\lambda_2((\phi_2-\phi_{2_0})^\dagger(\phi_2-\phi_{2_0}))^2\right) \\ & & \hspace{-0.25cm}
+\lambda_3(\phi_1^\dagger\phi_1)(\phi_2^\dagger\phi_2)+\lambda_4(\phi_1^\dagger\phi_2)(\phi_2^\dagger\phi_1)+\left[{m_{12}^2}(\phi_1^\dagger \phi_2) \right. +\dfrac{\lambda_5}{2}(\phi_1^\dagger\phi_2)^2 
\\ & & \hspace{6.05cm}
\left. +\lambda_6(\phi_1^\dagger\phi_1)(\phi_1^\dagger\phi_2)+ \lambda_7(\phi_2^\dagger\phi_2)(\phi_1^\dagger\phi_2)+\text{h.c.}\right],
\end{array}
\end{equation}
where
\begin{equation*}
{D_p}_\mu=\partial_\mu+\dot{\iota}\dfrac{g_p}{2}\sigma_i {W^i}_\mu+\dot{\iota}\dfrac{g_p'}{2}B_\mu, \qquad p=1 \text{ or } 2~,
\end{equation*}
\begin{equation*}
\phi_{i}=
\begin{bmatrix}
\phi^{+}_{i} \\
\eta_i + \dot{\iota}\chi_i +\nu_i \\
\end{bmatrix}
\text{ , \qquad}
\phi_{i_0}=
\begin{bmatrix}
0 \\
\tau_i \\
\end{bmatrix}	
\text{\qquad and \qquad}
\phi_{i}^{\dagger}=
\begin{bmatrix}
\phi^{-}_{i} & \eta_i - \dot{\iota}\chi_i +\nu_i
\end{bmatrix}.
\end{equation*}
The Higgs fields $\phi^{+}_{i}$, $\phi^{-}_{i}$, $\eta_i$ and $\chi_i$ are hermitian ($\phi^{\pm}_{i}$ are charged whereas the others are neutral), $\nu_i$ is the VeV of the doublet $\phi_i$, $\phi_{i_0}$ in the potential is the true minimum of the field $\phi_i$. 
The shape of the UDW-2HDM Higgs potential is
\begin{figure}[h!]
	\centering
	\includegraphics[scale=.4]{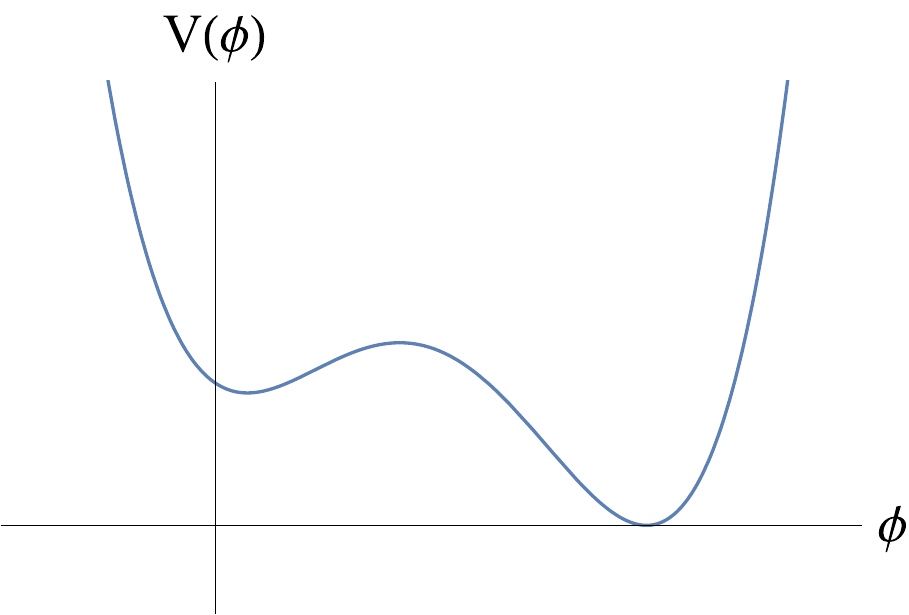}
	\caption{The uplifted double well potential}
	\label{fig:potential}
\end{figure}

The $Z_2$ symmetry ($\phi\rightarrow-\phi$) is very important in determining the stability of the Higgs fields.
There are two types of $Z_2$ symmetry breaking: $\left.1\right)$ soft; and $\left.2\right)$ hard. The term containing $m_{12}^{2}$ describes the soft $Z_2$ symmetry breaking, whereas the terms containing $\lambda_{6}$ and $\lambda_{7}$ describe the hard $Z_2$ symmetry breaking. In the absence of these terms along with no cross kinetic term i.e. $\chi=0$, the UDW-2HDM's Higgs Lagrangian has a perfect $Z_{2}$ symmetry. 

The extrema of the potential are found by taking
\begin{equation}\label{extremaconditions}
\dfrac{\partial V_H}{\partial \phi_1} \Big{|}_{\phi_1=\left\langle \phi_1\right\rangle }=\dfrac{\partial V_H}{\partial \phi_1^\dagger}\Big |_{\phi_1^\dagger= \left\langle \phi_1^\dagger\right\rangle}=0 
\text{\quad and \quad} 
\dfrac{\partial V_H}{\partial \phi_2}\Big |_{\phi_2=\left\langle \phi_2\right\rangle }=\dfrac{\partial V_H}{\partial \phi_2^\dagger}\Big |_{\phi_2^\dagger=\left\langle \phi_2^\dagger\right\rangle }=0.
\end{equation}
The most general solution of the conditions (\ref{extremaconditions}) is
\begin{equation*}\label{VeV}
\left\langle \phi_1\right\rangle =\frac{1}{\sqrt{2}}\begin{pmatrix}
0 \\ \nu_1
\end{pmatrix}
\text{\qquad and \qquad}
\left\langle \phi_2\right\rangle =\frac{1}{\sqrt{2}}\begin{pmatrix}
u \\ \nu_2
\end{pmatrix}.
\end{equation*}
We must remember that now $\nu^2=\nu_1^2+\left| \nu_2^2\right|+u^2$, where $\nu=1 / \sqrt[4]{2{G_F}^2}\approx246\text{GeV}$ is the VeV of the Higgs field in the SM.
For a nonzero $u$, the ``charged'' type dark energy is obtained, which obviously is not the case. For this reason, choose $u=0$. Solving eq. (\ref{extremaconditions}) for the potential truncated to the forth power of field gives
\begin{equation}\begin{array}{rcl}\label{1stcondition}
& & \rho_1\Lambda_1m_{11}^2(\nu_1-\tau_1)+\rho_1\Lambda_1^2m_{11}^4(\nu_1-\tau_1)^3+\rho_3\Lambda_3\lambda_1\nu_1(\nu_1^2-\tau_1^2)+{m_{12}^2}^{*}\nu_2 \\ && +(\lambda_3+\lambda_4+\lambda_5^{*})\nu_1\nu_2^2+(\lambda_6+2\lambda_6^{*})\nu_1^2\nu_2+\lambda_7^{*}\nu_2^3=0,
\end{array}
\end{equation}
\begin{equation}\begin{array}{rcl}\label{2ndcondition}
& & \rho_2\Lambda_2m_{22}^2(\nu_2-\tau_2)+\rho_2\Lambda_2^2m_{22}^4(\nu_2-\tau_2)^3+\rho_4\Lambda_4\lambda_2\nu_2(\nu_2^2-\tau_2^2)+{m_{12}^2}\nu_1 \\ && +(\lambda_3+\lambda_4+\lambda_5)\nu_1^2\nu_2+\lambda_6\nu_1^3+(2\lambda_7+\lambda_7^{*})\nu_1\nu_2^2=0.
\end{array}
\end{equation}
Imposing the $Z_2$ symmetry ($\chi=0 \text{,\quad} m_{12}^2=0 \text{,\quad} \lambda_6=0 \text{,\quad} \lambda_7=0$), which makes the lightest Higgs field is stable, on eqs. (\ref{1stcondition}, \ref{2ndcondition}) with $\lambda_3+\lambda_4+\lambda_5=0$ \cite{USMAN2015arXiv150607099U}\footnote{With $\lambda_3+\lambda_4+\lambda_5\neq0$ the solution to the VeV contain complex part} we get four solutions for $\nu_1$ and $\nu_2$, which are
\begin{eqnarray}
&\nu_1 = \tau_1
~, \text{ } &\nu_2 = \tau_2 ~; \label{electroweaksymmetricvacuum}
\\
&\nu_1 = \tau_1 ~, \text{ } &\nu_2 =\dfrac{2 m_{22}^4 \Lambda _2^2 \rho _2 \tau _2-\lambda _2 \Lambda _4 \rho _4 \tau _2-\sqrt{\Xi_2}}{2 \left(m_{22}^4 \Lambda _2^2 \rho _2+\lambda _2 \Lambda _4 \rho _4\right)} ~; \label{inertvacuum1}
\\
&\nu_1 =\dfrac{2 m_{11}^4 \Lambda _1^2 \rho _1 \tau _1-\lambda _1 \Lambda _3 \rho _3 \tau _1-\sqrt{\Xi_1}}{2 \left(m_{11}^4 \Lambda _1^2 \rho _1+\lambda _1 \Lambda _3 \rho _3\right)} ~, \text{\qquad} &\nu_2 = \tau_2 ~; \label{inertvacuum2}
\\
&\nu_1 =\dfrac{2 m_{11}^4 \Lambda _1^2 \rho _1 \tau _1-\lambda _1 \Lambda _3 \rho _3 \tau _1-\sqrt{\Xi_1}}{2 \left(m_{11}^4 \Lambda _1^2 \rho _1+\lambda _1 \Lambda _3 \rho _3\right)} ~, \text{\qquad} &\nu_2 =\dfrac{2 m_{22}^4 \Lambda _2^2 \rho _2 \tau _2-\lambda _2 \Lambda _4 \rho _4 \tau _2-\sqrt{\Xi_2}}{2 \left(m_{22}^4 \Lambda _2^2 \rho _2+\lambda _2 \Lambda _4 \rho _4\right)} ~; \label{mixedvacuum}
\end{eqnarray}
where
$$\Xi_1=-4 m_{11}^6 \Lambda _1^3 \rho _1^2-4 m_{11}^2
\lambda _1 \Lambda _1 \Lambda _3 \rho _1 \rho _3-8 m_{11}^4 \lambda _1 \Lambda _1^2 \Lambda _3 \rho _1 \rho _3 \tau _1^2+\lambda _1^2 \Lambda _3^2
\rho _3^2 \tau _1^2,$$
$$\Xi_2=-4 m_{22}^6 \Lambda
_2^3 \rho _2^2-4 m_{22}^2 \lambda _2 \Lambda _2 \Lambda _4 \rho _2 \rho _4-8 m_{22}^4 \lambda _2 \Lambda _2^2 \Lambda _4 \rho _2 \rho _4 \tau _2^2+\lambda
_2^2 \Lambda _4^2 \rho _4^2 \tau _2^2.$$

If we choose $\tau_2=0$ then the fields $\phi_1^\pm$ and $\chi_1$ become Goldstone bosons and the other fields become physical. With $\tau_2=0$, the Yukawa interactions are described by the interaction of $\phi_1$ with fermions (as $\phi_2$ does not couple to fermions but appears in loops). 
This scenario makes the lightest field of $\phi_2$ stable. 
\section{Higgs fields as dark energy}\label{sec:HiggsDE}
For the field $\phi_2$ to be the dark energy field, 
we need to solve the Euler-Lagrange equations in FRW flat, ($\kappa$), Universe, $(\sqrt{-g}=a(t)^3)$, 
which for the fields $\phi_{2}^\pm$, $\eta_2$ and $\chi_2$ are
\begin{equation}\begin{array}{rcl}\label{etaequationofmotion}
&& \hspace{-1cm} \ddot{\eta}_2 + 3\dfrac{\dot{a}}{a}\dot{\eta}_2 + \dfrac{1}{2}\eta_2\left(\nu^2\left(\lambda_3+\lambda_4+\lambda_5\right)+2 m_{22}^2\Lambda_2 \rho_2e^{m_{22}^2 \Lambda_2\left(\chi_2{}^2+\eta_2{}^2+2\phi_2^c{}^2\right) /2} \right. \\ && \hspace{-1cm} \left. +\lambda_2 \Lambda_4 \rho_4 e^{\lambda_2 \Lambda_4\left(\chi_2{}^2+\eta_2{}^2+2 \phi_2^c{}^2\right){}^2 /8} \left(\chi_2{}^2 + 2\phi_2^c{}^2\right)\right) + \dfrac{1}{2}\lambda_2\Lambda_4\rho_4e^{\lambda_2\Lambda_4\left(\chi_2{}^2+\eta_2{}^2+2 \phi_2^c{}^2\right){}^2 /8} \eta_2{}^3=0,
\end{array}
\end{equation}
\begin{equation}\begin{array}{rcl}\label{chiequationofmotion}
&& \hspace{-1cm}\ddot{\chi }_2+3 \dfrac{\dot{a}}{a}\dot{\chi }_2+\dfrac{1}{2}\chi _2 \left(\nu ^2\left(\lambda _3+\lambda _4-\lambda _5\right)+2 m_{22}^2
\Lambda _2 \rho _2e^{m_{22}^2 \Lambda _2\left(\chi _2{}^2+\eta _2{}^2+2 \phi _2^c{}^2\right) /2} \right.\\ && \hspace{-1cm} \left. +\lambda _2 \Lambda _4 \rho _4e^{\lambda _2 \Lambda _4\left(\chi _2{}^2+\eta _2{}^2+2 \phi _2^c{}^2\right){}^2 /8}\left(\eta _2{}^2+2\phi _2^c{}^2\right)\right)+\dfrac{1}{2}\lambda _2
\Lambda _4 \rho _4e^{\lambda _2 \Lambda _4\left(\chi _2{}^2+\eta _2{}^2+2 \phi _2^c{}^2\right){}^2 /8}\chi _2{}^3=0,
\end{array}
\end{equation}
\begin{equation}\begin{array}{rcl}\label{chargedHiggsequationofmotion}
\ddot{\phi }_2^c+3 \dfrac{\dot{a}}{a}\dot{\phi }_2^c &+& \phi _2^c \left(\nu ^2 \lambda _3+2 m_{22}^2 \Lambda _2 \rho _2e^{m_{22}^2 \Lambda _2\left(\chi _2{}^2+\eta _2{}^2+2 \phi _2^c{}^2\right) /2} \right. \\ &+& \left. \lambda_2\Lambda_4\rho_4e^{\lambda_2\Lambda_4\left(\chi_2{}^2+\eta_2{}^2+2\phi_2^c{}^2\right){}^2 /8} \left(\chi_2{}^2+\eta_2{}^2+2\phi_2^c{}^2\right)\right)=0,
\end{array}
\end{equation}		
where $c$ is $+$ or $-$.
The energy density and pressure after expansion of UDW-2HDM Higgs Lagrangian for physical fields become
\begin{equation}\begin{array}{rcl}\label{rhoPeta2}
\rho _{\text{DE}}/\text{P}_{\text{DE}}= && \dfrac{1}{2}\dot{\chi }_2{}^2+\dfrac{1}{2} \dot{\eta }_2{}^2+\dfrac{1}{2}\dot{\phi }_2^c{}^2\pm \left(\rho _1+\rho _3 +\dfrac{1}{2} \nu ^2\lambda _3\phi _2^c{}^2+\dfrac{1}{4}\nu ^2\left(\lambda _3+\lambda _4+\lambda _5\right)\eta _2{}^2 \right. \\ && +\dfrac{1}{4}\nu ^2\left(\lambda _3+\lambda _4-\lambda _5\right)\chi _2{}^2+\rho _2e^{m_{22}^2 \Lambda _2\left(\chi _2{}^2+\eta _2{}^2+2\phi _2^c{}^2\right) /2} \\ && \left.+\rho _4e^{\lambda _2 \Lambda _4\left(\chi _2{}^4+2 \chi _2{}^2 \eta _2{}^2 + \eta _2{}^4 +4\chi _2{}^2 \phi _2^c{}^2 +4 \eta _2{}^2 \phi _2^c{}^2+4\phi _2^c{}^4\right) /8}\right).
\end{array}
\end{equation}

The initial conditions used are ${\eta_2}_{ini}=M_P$, ${\chi_2}_{ini}=M_P$, ${\phi_2^c}_{ini}=0$, ${\dot{\chi_2}}_{ini}=0$, ${\dot{\chi_2}}_{ini}=0$, ${\dot{\phi_2^c}}_{ini}=0$.
The masses of the Higgs bosons in the analysis are taken to be $m_{\eta_2}=m_{\chi_2}=1.0247\times 10^{-59}\text{ GeV}$, the charged Higgs mass is arbitrary.
%
The solution of the eqs. (\ref{etaequationofmotion}, \ref{chiequationofmotion}, \ref{chargedHiggsequationofmotion}) along with Friedmann equations is shown below in the graphs.

\begin{figure}[h!]
	\hspace{-1.0cm}
	\includegraphics[scale=.5]{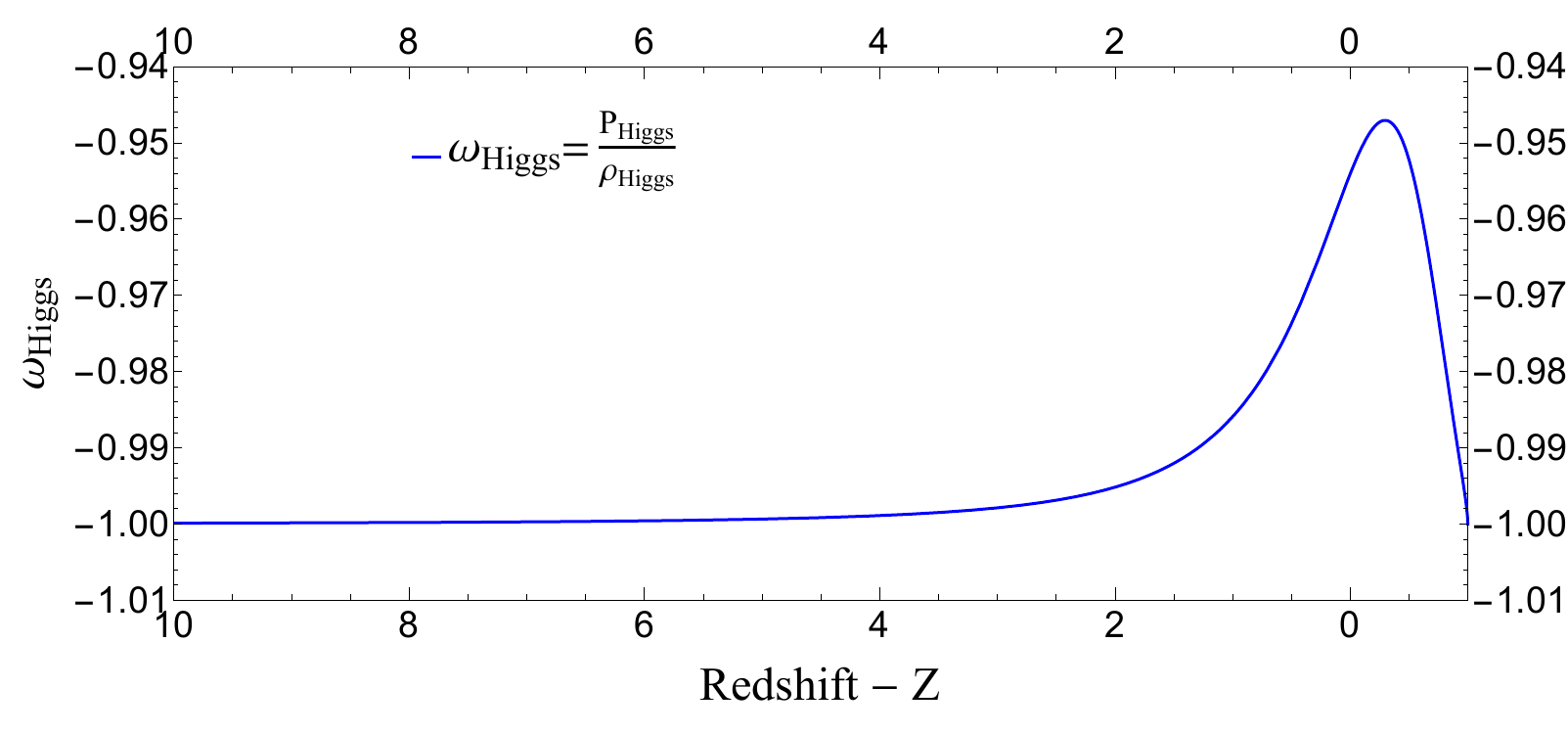}
	\caption{Effective equation of state parameter for Higgs fields $\omega_{Higgs}$, as seen it starts with $-1$ then evolves towards quintessence regime after large enough time it comes back at $-1$.}
	\label{fig:OmegaHiggs}
\end{figure}
During the initial stages $Z\gg1$, the evolution of the Higgs fields, $\eta_2$, $\chi_2$ and $\phi_2^{\pm}$, is frozen, 
and acts as a negligibly small vacuum energy component with $\omega=-1$. As time proceeds the Higgs fields begin to evolve towards the minimum of the potential, the energy density in the Higgs fields starts to dominate cosmologically. During the evolution, $\omega_{Higgs}$ starts to increase and becomes $>-1$ as shown in fig. (\ref{fig:OmegaHiggs}). In the very late (future) Universe ($Z\ll0$), the fields come to rest at the minimum of the potential and a period with $\omega=-1$ is re-achieved to give an accelerating Universe similar to a pure cosmological constant. 

As discussed before, the Higgs field stability is provided by imposing $Z_2$ symmetry. The lightest Higgs fields, $\eta_2$ and $\chi_2$, do not decay into any other Higgs field (since these fields are lighter than the SM-like and charged Higgs) or into fermions (since they do not couple to them at tree level).
Note that the initial conditions for the charged field took the dark energy (vacuum) not to be charged. Since it does not contribute to relic density, this does not mean that it can not exist, it can appear in the loop process.
\section{Conclusion}\label{sec:conclusion}
In the work presented, we assumed that dark energy is actually some scalar field which is present as the Higgs in a model where the potential has the non-degenerate vacua, we called this model uplifted double well two-Higgs doublet model (UDW-2HDM).

We found that if the present Universe is described by the true vacua of UDW-2HDM then the component fields of the second doublet $\phi_2$ (which acts as the inert doublet) can be one possible candidate for the dark energy. As the present contribution of the dark energy to the critical energy density is about $0.7$, this value is obtained by taking the mass of the CP-even field's mass small ($O(10^{-59})$GeV). The most important thing is that with the initial conditions set, the mass of the charged ($\phi_2^\pm$) field becomes arbitrary. Hence this model will fit for any value of mass of $\phi_2^\pm$. One also needs to keep in mind that the values of masses were chosen arbitrarily so as to get dark energy relic density $\approx 0.7$. Changing the values of the masses, the relic density does not change much. 

It should also be mentioned that if we remove the $Z_2$ symmetry, the second Higgs doublet does not remain inert. Thus in the case of $Z_2$ violation (soft or hard), the CP-even Higgs fields will mix by an angle $\beta$. In that case a new parameter ($\beta$) will arise in the theory. Obtaining a dark energy candidate in that model will require fine tuning in the Yukawa interactions in such a way that either the dark energy field does not couple or couple very weakly with the fermions.
\bibliographystyle{ws-procs961x669}
\nocite{*}
\bibliography{Muhammad-Usman.bib}
\end{document}